\documentstyle[11pt]{article}
\newcommand{\noi}{\noindent}
\newcommand{\beq}{\begin{equation}}
\newcommand{\eeq}{\end{equation}}
\newcommand{\beqr}{\begin{eqnarray}}
\newcommand{\eeqr}{\end{eqnarray}}

\newcommand{\bl}{\vspace{.2in}}
\begin{document}
{}~\hfill{hep-th/yymmnn}\\
{}~\hfill{UICHEP-TH/94-5}\\
{}~\hspace*{\fill}{\today}\\

\vspace{.6in}
\centerline {\LARGE\bf Quantum Mechanics of Multi-Prong Potentials}
\vspace{.5in}

\centerline {A. Gangopadhyaya$^{(1)}$, A. Pagnamenta$^{(2)}$ and
U. Sukhatme$^{(2)}$}
\vspace{0.3in}
\centerline{$^{(1)}$Department of Physics, Loyola University Chicago, IL
60626}
\vspace{0.15in}
\centerline{$^{(2)}$Department of Physics, University of Illinois at
Chicago,}
\centerline{845 W. Taylor Street, Chicago, Illinois 60607-7059}
\vspace {1in}

\centerline{\bf Abstract}
\vspace{0.3in}

We describe the bound state and scattering properties of a quantum
mechanical particle in a scalar $N$-prong potential. Such a study is of
special interest since these situations are intermediate between one and two
dimensions. The energy levels for the special case of
$N$ identical prongs exhibit an alternating pattern of non-degeneracy and
$(N-1)$ fold degeneracy. It is shown that the techniques of
supersymmetric quantum mechanics can be used to generate new solutions.
Solutions for prongs of arbitrary lengths are developed.
Discussions of tunneling in $N$-well potentials and of scattering for
piecewise constant potentials are given. Since our treatment is for
general values of $N$, the results can be studied in the large $N$ limit.
A somewhat surprising result is that a
free particle incident on an $N$-prong vertex undergoes continuously increased
backscattering as the number of prongs is increased.

\newpage

\section{Introduction}

\hspace{.25in}  Various aspects of the solutions of the Schr\"{o}dinger
equation for both scalar and vector potentials on a wide variety of networks
have been discussed by several authors
[1-8].
Recent investigations have been  motivated in part by the considerable
interest in mesoscopic
systems and the experimental observation of persistent
currents\cite{Levy90}. Most of the
results correspond to piecewise constant
scalar potentials $V=V_0$ \cite{Rammal82,Shapiro83,Porod93},
vector potentials
{\bf A} associated with uniform magnetic fields \cite{Buttiker83}, or
$\delta-$function potentials at the network
vertices \cite{Weger73,Avron88,Tekman93}. In this paper we discuss the
bound states and the scattering properties of a single particle
moving in an arbitrary scalar $N$-prong potential. The treatment is
kept simple in order to clearly
show the generalization from the two-prong case, which is just the
familiar one-dimensional (particle on a line) problem. The properties
of the eigenstates of $N$-prong potentials are not a priori evident,
since this system is in some sense intermediate between one and two
dimensions.\bl

The plan of this paper is as follows. In Sec. 2, we define
$N$-prong potentials and the boundary conditions which the wave function
must satisfy. The normalization of wave functions
and their orthogonality properties are also discussed.
Sec. 3 contains a discussion of
bound states for potentials with $N$ identical prongs. The general
solution
is quite easy to obtain and is in fact closely related to the solution
of a symmetric one-dimensional potential.
Sec. 4 contains a discussion of bound states in $N$-prong potentials
with non-identical prongs. Several analytical and
numerical solutions are given to
illustrate eigenfunction properties. In particular, we show that a
generalized version of the usual rule about one extra node
for each higher eigenstate
is probably true but the theorem about non-degeneracy in
one-dimensional problems is not. Since the concept of
supersymmetry has yielded many interesting results for
one-dimensional quantum mechanics\cite{Witten81,Cooper94}, it is natural
to investigate what new results come from
using supersymmetry for $N$-prong potentials. This is discussed in Sec. 5.
Sec. 6 illustrates the use of lowest order perturbation theory to obtain
bound state energies. Scattering is discussed in Sec. 7. For the special case
of no potential in any of the $N$ prongs, very interesting properties
are obtained which are quite different from naive expectations. Finally, in
Sec. 8, we extend the usual discussion of tunneling for a double well
potential to the case of $N$-well potential systems.

\section{Multi-Prong Potentials}

\hspace{.25in}   We are addressing the problem of formulating and solving the
Schr\"{o}dinger equation  for a single particle constrained to
remain in a space made up of $N$ lines meeting at a vertex point
(Fig. 1). The prongs are labeled by the indices $i$
$(i=1,2,\cdots,N)$ and the position on any prong is given by the
positive coordinate $x_i$, with $x_i=0$ being the vertex point. The
potential is fully specified by giving the scalar potentials
$V_i(x_i)$ on each prong $i$. The overall wave function
$\psi(\vec{x})$ is composed of the individual wave functions
$\psi_i(x_i)$ on each prong $i$:
\beq
\psi(\vec{x}) \equiv \{ \psi_1(x_1), \cdots,
\psi_N(x_N)\}.\label{def}
\eeq
We are using the arrow notation to indicate the appropriate
$N$-tuple. The physical requirement of single-valuedness of
the wave function at the vertex implies for the component
wave functions
\beq
\psi_1(0)=\psi_2(0)=\cdots =\psi_N(0).\label{BC1}
\eeq
In this paper, we will not be considering $\delta$-function potentials
at the vertex. Consequently, the second
vertex condition requires the sum of all
derivatives to add up to zero \cite{Coulson54,Avron88}:
\beq
\left.\sum_{i=1}^{N} \frac{d\psi_i}{dx_i}\right|_{x_i=0}
=0.\label{BC2}
\eeq
This can be readily derived as follows. The time independent
Schr\"{o}dinger equation, reads $\nabla^2\psi=(V-E)\psi$,
taking units in which $\hbar=2m=1$.
We assume that the potential is finite in
the neighborhood of the vertex, and integrate over a small sphere
of radius $\epsilon$ with the vertex as the center. Using the
divergence theorem and letting $\epsilon \rightarrow 0$ yields
$\lim_{\epsilon \rightarrow 0}\int_A \nabla\psi \cdot dA=0$, which
reduces to Eq. (\ref{BC2}) for an $N$-prong potential. Note that for
the special case $N=2$, one has the standard one-dimensional
situation and, taking into account the relative directions of
$x_1$ and $x_2$,  Eq. (\ref{BC2}) is just the statement that the
derivative $d\psi/dx$ is continuous across the vertex.\bl

For the $N$-prong system, we  call two functions
$\psi(\vec{x})$ and $\phi(\vec{x})$ orthogonal if
\beq
\sum_{i=1}^{N} \int \psi_i^*(x_i) \phi_i(x_i) dx_i = 0,
\eeq
where the sum indicates that all prongs are included and each
integral runs
over the appropriate prong length.
The normalization condition, which also includes all prongs, reads
\beq
\sum_{i=1}^{N} \int \psi_i^*(x_i) \psi_i(x_i) dx_i = 1.
\eeq
\bl

The eigenfunctions of the Schr\"odinger equation
in a multi-prong potential not only have to satisfy the vertex
conditions given
in Eqs. (\ref{BC1}) and (\ref{BC2}), but also appropriate boundary
conditions at
each prong end.
These conditions depend on the energy $E$ and the
specific behavior of $V_i(x_i)$ for the maximum allowed
value of $x_i$.
Eigenfunctions corresponding to bound state solutions have to
vanish at the prong ends.
For potentials which reach a constant value fast enough at the prong ends,
one has standard plane
wave solutions $(e^{ik{x_i}},e^{-ik{x_i}})$ for the scattering states.

\section{Bound States: Identical Prongs}

\hspace{.25in}   We begin by discussing the interesting special case of $N$
identical
prongs, i.e.
$V_i(x_i)=V(x_i).$ Note that this does not imply that the wave
functions are
identical on all the prongs. In fact, Eq.
(\ref{BC2}) implies the wave functions are often different, since their
derivatives at the vertex have to
add up to zero. This statement is the generalization of
the two-prong case, where one has symmetric and antisymmetric
solutions. \bl

The general bound state solution for $N$ identical prongs is easily
obtained
from the two-prong situation. For two identical prongs, one has the
situation
schematically shown in Fig. 2.
We can define the single variable $x$ such that
$x=x_1$ for positive $x$, and $x=-x_2$ for negative $x$.
Effectively, one is
mapping the two prongs onto the real axis, $ -\infty < x < \infty.$
Clearly,
this is the familiar situation of a symmetric, one-dimensional
potential. Its
eigenstates correspond to even and odd solutions $\psi^{(n)}(x)$ at
energies
$E_n$, labeled by a quantum number $n$ ($n=0,\; 1, \;2,\cdots$).
\bl

\noindent{\bf Theorem:}
{}~The eigenstates of a potential with $N$ identical prongs
can be constructed from the eigenfunctions $\psi^{(n)}(x)$  of the
corresponding symmetric two-prong system,
and have exactly the same eigenenergies. Explicitly,
\begin{equation}
\psi^{(n)}(\vec{a},\vec{x}) \equiv
\left\{ a_1 \psi^{(n)}(x_1), a_2 \psi^{(n)}(x_2), \cdots ,
a_N \psi^{(n)}(x_N) \right\}, \;\;\;\;(x_i>0).\label{eigfct}
\end{equation}
where ${\vec{a}}$ is a compact notation for the $N$-tuple
$(a_1,a_2,\cdots,a_N)$.
For even numbered states
the boundary conditions at the vertex imply
$a_1=a_2=\cdots=a_N$; for odd numbered states one has the
constraint
$\sum_ia_i=0$, and there is an $(N-1)$ fold degeneracy. \bl

\noindent
{\bf Proof:} By construction, the wave function
$\psi^{(n)}(\vec{a},\vec{x})$ for an
eigenenergy $E_n$ satisfies all the boundary conditions at the
prong ends. It
only remains to show that the vertex conditions, Eqs. (\ref{BC1})
and (\ref{BC2})
are also satisfied by Eq. (\ref{eigfct}).
a) For even eigenfunctions of a two-prong potential one has
$\psi'(0)=0,$ and in general $\psi(0)\neq 0.$ Eq. (\ref{BC2})
 is therefore trivially
satisfied, and Eq. (\ref{BC1}) implies $a_1=a_2=\cdots=a_N$. The
eigenfunction is therefore determined upto one overall normalization
constant, and the energy level is non-degenerate.
b) For odd eigenfunctions of a two-prong potential, one has
$\psi'(0)\neq 0,$ and  $\psi(0)=0$. Now, Eq. (\ref{BC1}) is
trivially satisfied
(since each wave function vanishes at the vertex) and the
derivative
condition Eq. (\ref{BC2}) implies $\sum_ia_i=0$, as claimed. This
constraint is not sufficient to fully determine the constants $a_i$,
and the levels have an $(N-1)$ fold degeneracy.
The completeness of states in Eq. (\ref{eigfct}) seems intuitively
reasonable, and can be established from the completeness of the
eigenfunctions of a symmetric one dimensional potential.
\bl

As a first illustrative example, consider a particle of mass $m$
moving in a harmonic potential with three identical prongs
\beq
V(x_i) = \frac{1}{2}m \omega^2 x_i^2.
\eeq
Each prong corresponds to the positive half of a harmonic
oscillator potential. The ground state of this
three-prong potential will have the
same energy $E_0={1 \over 2} \hbar \omega$ as the two-prong case. The
un-normalized eigenfunction is given by
\begin{equation}
\psi^{(0)}(\vec{a},\vec{x}) \equiv
\left\{ a\psi^{(0)}(x_1),a\psi^{(0)}(x_2), a\psi^{(0)}(x_3) \right\},
\label{wave_ho}
\end{equation}
where, $\psi^{(0)} \propto e^{-\omega{x^2}/4}$ is the
Gaussian harmonic oscillator ground state wave function.
We assume $\psi^{(n)}(x)$ to be normalized to
unity over one prong, so that the wave function in Eq. (\ref{eigfct})
is properly normalized.
The next level at energy $E_1={3 \over 2} \hbar \omega$
is doubly degenerate.
Eigenfunctions at this level are given by
\begin{equation}
\psi^{(1)}(\vec{a},\vec{x})
\equiv
\left\{ a_1 \psi^{(1)}(x_1), a_2 \psi^{(1)}(x_2), a_3 \psi^{(1)}(x_3) \right\},
\label{wave_ho2}
\end{equation}
where, $\psi^{(1)} \propto xe^{-\omega{x^2}/4}$ is the wave function for
the first excited state of the
two-prong case. There is an infinite number of
wave functions
$\psi^{(1)}(\vec{a},\vec{x})$ that correspond to the same energy
${3 \over 2} \hbar \omega$. Each
of these wave functions
is characterized  by a set of coefficients $a_i$ obeying the
constraint
$\sum_i a_i=0$. Due to this constraint, we see that all allowed
$a_i$ span a
two-dimensional space, thus leading to a two fold degeneracy at
this level.
The following two functions then provide one choice for an orthonormal
basis in this two-dimensional, two-fold degenerate space:
\beqr
\psi^{(1)} (\vec{a}_1; \vec{x}), && \vec{a}_1 \equiv
\left( {1\over \sqrt{2}},0, {{-1} \over \sqrt{2}}\right) ;
\nonumber\\
&& \label{excited_ho}\\
\psi^{(1)} (\vec{a}_2; \vec{x}), && \vec{a}_2 \equiv
\left( {1\over \sqrt{6}}, {{-2} \over \sqrt{6}}, {1\over \sqrt{6}} \right).
\nonumber
\eeqr
This pattern of nondegeneracy/degeneracy keeps repeating as we
consider higher energy states. All even numbered states are
nondegenerate and all odd numbered states have two fold degeneracy.
The three lowest eigenstates are shown in Fig. 3. The normalization
has been chosen such that
$\int_0^\infty \psi^{(n)*}(x) \psi^{(n)}(x) dx =1.$  \bl

Similarly, the eigenfunctions for the potential $V(x_i)=0$,~~ $x_i\leq 1$ are
shown by the dashed lines in Fig. 4. The wave functions are sinusoidal, and
the energies are $E_n=n^2 \pi^2/4~~ (n=1,2,\cdots)$, since they
correspond to a symmetric two-prong potential which is an infinite square
well of width 2. \bl

{}From the above examples, it becomes clear that a potential with $N$ identical
prongs will have alternate nondegenerate and $(N-1)$ fold degenerate energy
levels. Clearly, the familiar  non-degeneracy property of one-dimensional
potentials is maintained for $N=2$, but not for situations with more prongs.

\section{Bound States: Non-Identical Prongs}

\hspace{.25in}     Let us now consider the general case of an $N$-prong
system with potential $V_i(x_i)$ on the $i$-th prong.
Let $\psi_i(x_i)$ be the solution of the Schr\"{o}dinger equation with
energy $E$ along
that  prong:
$$-{ {d^2\psi_i} \over {dx_i^2}	} + \left( V_i-E\right) \psi_i=0,
\;\;\;\;(i=1,2,\cdots,N).$$
Of the two linearly independent solutions to this equation, let
$\phi_i(x_i,E)$ be the one that vanishes at the end point of the  $i$-th prong.
This implies $\psi_i(x_i,E)=a_i \phi_i(x_i,E)$, where $a_i$ is a constant.
These $\psi_i(x_i,E)$, glued together properly so as to satisfy vertex
requirements, will produce the wave function
for the entire domain. From vertex conditions (\ref{BC1}) and (\ref{BC2}),
we get
\beq \label{BC1S}
\psi_i(0, E)=a_i \phi_i(0,E) \equiv {U},
\eeq
and
\beq \label{BC2S}
\sum_{i=1}^N \psi_i'(0, E) = \sum_{i=1}^N a_i \phi_i'(0,E)=0.
\eeq
The energy eigenvalues $E$ are then determined by eliminating the unknowns
$a_i$ and $U$ from the above set of $(N+1)$ linear equations. The eigenvalue
condition is:
\beq \label{ev}
{\rm det} \left|
\begin{array}{ccccc}
\phi_1(0,E) 	&0		&\cdots	&0		&1\\
0		&\phi_2(0,E) 	&\cdots	&0		&1\\
\cdots		&\cdots		&\cdots		&\cdots		&\cdots	\\
0		&0		&\cdots	&\phi_N(0,E)    &1\\
\phi_1'(0,E) 	&\phi_2'(0,E)   &\cdots &\phi_N'(0,E) 	&0
\end{array}
\right|=0.
\eeq
For any given eigenvalue $E_n$, Eq. (\ref{BC1S}) then determines all the
constants $a_i$ in terms of one unknown constant $U$. The corresponding
bound state wave function is:
\beq \label{psi}
\psi(\vec {x},E_n)=U\left\{\frac {\phi_1(x_1,E_n)}{\phi_1(0,E_n)},\cdots,\frac
 {\phi_N(x_N,E_n)}{\phi_N(0,E_n)}\right\}.
\eeq
If desired, the
constant $U$ can also be fixed by requiring overall normalization of
the wave function. \bl

To illustrate the procedure for determining eigenvalues, consider the
example of a particle that is free to move inside
a domain of three prongs of lengths $l_1$, $l_2$ and $l_3$. Since the
potential along the prongs is zero, the wave functions along them are
given by sinusoidal functions that vanish at the end points. An
eigenfunction of energy $E=k^2$ has the form:
\begin{equation}
\psi(\vec{a}, \vec{x})=\left\{
a_1 \sin k(l_1-x_1),
a_2 \sin k(l_2-x_2),
a_3 \sin k(l_3-x_3)
\right\}.
\label{free_wave}
\end{equation}
In this case,
Eq. (\ref{BC1S}) is:
\beq
a_1 \sin kl_1 = a_2 \sin kl_2 = a_3 \sin kl_3=U ,
\label{BC1'}
\eeq
where $U$ is the common value of the wave function at the origin.
The derivative condition (\ref{BC2S})
gives:
\beq
a_1 \cos kl_1 + a_2 \cos kl_2 + a_3 \cos kl_3=0.
\label{BC2'}
\eeq
The eigenvalue condition (\ref{ev}), when simplified reads
\beq
\cos kl_1\,\sin kl_2\,\sin kl_3+\sin kl_1\,\cos kl_2\,\sin kl_3+
\sin kl_1\,\sin kl_2\cos kl_3
=0.
\label{permute}
\eeq

As a specific case, consider the situation $l_2=0.8$ and $l_1=l_3=1$. For this
case, the solutions to Eq. (\ref{permute}) are found to be
$k=1.68, \pi, 3.61, 5.08, 2\pi, 7.17, 8.54, 3\pi,...$.
The solutions $k=m \pi ~ (m=1,2,...) $
are a consequence of maintaining partial symmetry by
taking two prongs to be identical $(l_1=l_3=1)$. The
corresponding wave functions are shown by the solid lines in Fig. 4. [For
comparison, we have also plotted the dashed lines corresponding to the wave
functions for the case of all three identical prongs of length 1]. From the
figure, it is apparent that the
ground state wave function at $E_0=1.68^2~~$ has no nodes.
The first excited state
at $E_1=\pi^2~$ has one node at the vertex point. Note that for this state,
the wave function in prong 2 is zero. The second excited state is at energy
$E_2=3.61^2~~$ and has two nodes, one in prong 1 and the other in prong 3.
This result is very suggestive - one expects one extra node to appear for
each higher eigenstate, similar to the familiar one-dimensional situation.

We have also studied the variation of the eigenvalues $k$ systematically as the
prong length $l_2$ is varied. The results are shown in Fig. 5. As discussed
above, the solutions $k=m\pi$ are present, and when $l_2$ is an
integral multiple of
$l_1 (=l_3)$, one has the interesting occurrence of
degeneracy and level crossing. The curves in Fig. 5 are labeled by the number
of nodes in the wave functions. At any fixed value of $l_2$, the number of
nodes increases with energy. Note that as $l_2 \rightarrow 0$, the
eigenvalues become doubly degenerate at $k=m\pi,~~(m=1,2,...)$. This is
intuitively clear since the limit $l_2 \rightarrow 0$ forces the wave
function to vanish at the vertex, thereby effectively breaking the
problem into two infinite square well potentials of widths $l_1=1$ and $l_3=1$.
Finally, in Fig. 6, we take $l_1 \ne l_3$, and plot eigenvalues $k$. This is a
completely asymmetric situation, and there is now no degeneracy
of energy levels. Again note that the number of nodes in the overall
wave function increases by one with increasing energy levels. \bl

As a second example, we determine the eigenstates of a three-prong
potential $V_i(x_i)={1 \over 4} \omega_i^2 x_i^2$, composed of
three harmonic oscillators of different
angular frequencies $\omega_i~~ (i=1,2,3)$,
and $\hbar=2m=1$. The wave functions which vanish at $x_i \rightarrow \infty$
are \cite{Merzbacher61}
\beq
\psi_i(x_i)=a_iD_{\nu_i}(\sqrt {\omega_i}x_i)~,~~\nu_i=\frac {E}{\omega_i}
- \frac{1}{2}~~,
\eeq
where $D_{\nu}$ is a parabolic cylinder function. The eigenvalue
condition [Eq. (\ref{ev})] gives
\beq
\sqrt {\omega_1}D_{\nu_1}'(0)D_{\nu_2}(0)D_{\nu_3}(0)
+ {\rm cyclic~~ permutations}~=~0~.
\eeq
The parabolic cylinder function and its derivative at the origin have
simple expressions in terms of gamma functions \cite {Merzbacher61}:
\beq
D_{\nu}(0)=2^{\frac {\nu}{2}} \frac {\Gamma( \frac {1}{2})}
{\Gamma({1 \over 2}-{\nu \over 2})}~~,~~
D_{\nu}'(0)=2^{{\nu \over 2} - {1 \over 2}} \frac {\Gamma(- {1 \over 2})}
{\Gamma(-{\nu \over 2})}~~.
\eeq
The solutions of the eigenvalue equation are then easily found. For the
choice $\omega_1=  1.0,~ \omega_2=  2.0,~ \omega_3=  3.0$,
one gets the four lowest eigenstates at energies $E=0.83, 1,94, 3.29, 3.85$.
There is no degeneracy since all $\omega$'s are different.
\bl

In the above two examples, exact analytic forms for the solutions
$\psi_i(x_i)$ were available. Even if this is not the case, it is easy to
use numerical Runge-Kutta techniques applied to each prong.

\section{Supersymmetric Quantum Mechanics on Multi-Pronged Domains}

\hspace{.25in}    Given any one-dimensional potential $V(x)$, the
powerful techniques of supersymmetric quantum mechanics can be used to generate
a partner potential $\tilde {V}(x)$ with the same
eigenvalues \cite{Witten81,Cooper94}. This property
has been extensively used to get a deeper understanding of exactly
solvable potentials and for generating improved approximation methods
(SWKB method \cite{Comtet85}, large $N$ method \cite{Imbo85}, etc.)
for determining eigenvalues. Supersymmetric quantum mechanics provides
an elegant formalism which includes and goes substantially beyond the
method of factorization previously applied to some
potentials \cite{Schrodinger40}. In this section, we show how the
ideas of supersymmetric quantum mechanics can be applied to a given
$N$-prong system $V(\vec{x})$ in order to generate solutions for a new
$N$-prong potential $\tilde{V}(\vec{x})$.

As in the previous section, consider a scalar $N$-prong potential
$V(\vec{x})$ composed of potentials $V_i(x_i)$ on prong $i~~(i=1,2,...,N)$.
Its eigenvalues and eigenfunctions are then given by
Eqs. (\ref{ev}) and (\ref{psi})
respectively. The un-normalized ground state wave function is
\beq
\psi^{(0)}(\vec{x},E_0) \propto \left\{\frac {\phi_1^{(0)}(x_1,E_0)}
{\phi_1^{(0)}(0,E_0)},\cdots,\frac
 {\phi_N^{(0)}(x_N,E_0)}{\phi_N^{(0)}(0,E_0)}\right\}.
\eeq
This can be used to define the superpotential $W(\vec{x})$ whose value
on prong $i$ is
\beq
W_i(x_i)=-{ {\phi_i^{(0)'}(x_i)} \over {\phi_i^{(0)}(x_i)} }.
\eeq
It is easy to check that
\beq
V_i(x_i)=W_i^2(x_i)-W_i'(x_i)+E_0.
\eeq
The supersymmetric partner potential is given by
\beq
\tilde {V_i}(x_i)=W_i^2(x_i)+W_i'(x_i)+E_0.
\label{V-tilde}
\eeq
The potentials $V_i(x_i)$ taken together make up the full $N$-prong
potential $\tilde {V}(\vec{x})$. From supersymmetric quantum mechanics,
we know that the solution of the Schr\"{o}dinger equation for potential
$\tilde V_i(x_i)$ and energy E is given by \cite{Cooper94}
\beq \label{n}
\tilde \phi_i(x_i,E) = a_i \left( -{d \over {dx_i} }+
{       {\phi^{(0)'}_i(x_i)} \over {\phi^{(0)}_i(x_i)} }
\right) \phi_i(x_i,E),
\label{tilde_phi}
\eeq
where $a_i$ are constants. Since $\phi_i(x_i,E)$
vanishes at the prong ends, so do all the $\tilde \phi_i(x_i,E)$. At the
vertex, one wants
\beq \label{nn}
\tilde \phi_i(0,E)=\tilde \phi_j(0,E)\;\;\;{\rm and}\;\;\;
\sum_i \tilde\phi_i'(0,E)=0.
\eeq
This gives an eigenvalue condition similar to Eq. (\ref{ev}).

\beq \label{tildeev}
{\rm det} \left|
\begin{array}{ccccc}
\tilde\phi_1(0,E) 	&0		&\cdots	&0		&1\\
0		&\tilde{\phi}_2(0,E) 	&\cdots	&0		&1\\
\cdots		&\cdots		&\cdots		&\cdots		&\cdots	\\
0		&0		&\cdots	&\tilde{\phi}_N(0,E)    &1\\
\tilde{\phi}_1'(0,E) 	&\tilde\phi_2'(0,E)   &\cdots &\tilde\phi_N'(0,E) &0
\end{array}
\right|=0.
\eeq
\bl

In general, the eigenvalues obtained from Eq. (\ref{tildeev}) will be
different from those coming from Eq. (\ref{ev}). However, there are two
important situations where the {\it same} eigenvalues result. This happens
for the special case of two prongs ($N=2$) and for the case of all
identical prongs. These results can be understood from a physical
viewpoint, since $N=2$ is the
standard one-dimensional situation treated in supersymmetric quantum
mechanics, and furthermore, as we have seen in Sec. 3,
these are the same eigenvalues for the identical prong case.
Thus we see the machinery of supersymmetric quantum mechanics can be
immediately applied to get eigenvalues and eigenfunctions of the partner
potential $\tilde V(\vec {x})$
for the identical prong case. The equality of eigenvalues from Eqs.
(\ref{ev}) and (\ref{tildeev}) can also be established from a mathematical
viewpoint using Eqs. (\ref{n}) and (\ref{nn}). \bl

\section{Perturbation Theory}

\hspace{.25in}  Having solved a multi-prong problem, it
is of interest to see the influence of a small perturbation of the
potential in one
of the prongs, say, on the energy levels.
Here, for simplicity, we will only deal with an unperturbed
system with three identical prongs.
Consider an unperturbed system with potential $V_i(x_i)=V_0(x_i)$ on each
prong. Its completely symmetric ground state
\begin{equation} \label{mm}
\psi^{(0)}(\vec{1},\vec{x}) \equiv
{1\over{\sqrt{3}}} \left\{\psi^{(0)}(x_1), \psi^{(0)}(x_2),
\psi^{(0)}(x_3) \right\},
\nonumber
\end{equation}
is described in terms of $\psi^{(0)}(x)$; the  ground state
eigenfunction of a
two-prong system with the same potential $V_0(x_i)$ in each prong.
We have chosen $\psi^{(0)}(x)$ to be normalized to
unity over one prong, so that the wave function in Eq. (\ref {mm})
is properly normalized.
The eigenenergy of this state is also the same as the
ground state
energy of the two-prong system.
The first excited state for this three-prong system is doubly
degenerate. We
choose the following two orthonormal states as our base states:
\begin{equation}
\psi^{(1)}(\vec{a}_1;\vec{x}) \equiv
{1\over{\sqrt{2}}}\left\{\psi^{(1)}(x_1),0,-\psi^{(1)}(x_3) \right\},
\label{psi1}
\end{equation}
and
\begin{equation}
\psi^{(1)}(\vec{a}_2;\vec{x}) \equiv
{1\over{\sqrt{6}}}\left\{\psi^{(1)}(x_1), -2\psi^{(1)}(x_2),
\psi^{(1)}(x_3) \right\}.
\label{psi2}
\end{equation}
Both of these states have exactly the same energy as the first
excited
state of the two-prong system. \bl

Now, let us include a perturbation $V_I$ that
is nonvanishing in only one prong. The energy
of the system is shifted and the degeneracy of the states of
Eqs. (\ref{psi1}) and (\ref{psi2}) is lifted.
Interestingly, only one of the eigenvalues of the first
excited state
changes, while the other one remains the same, thus causing the
split.
We compute here such shifts for the ground state as well the first excited
state
of the system. For numerical concreteness only, in the following
example,
we use the harmonic oscillator potential for $V_0$ and $\alpha x_1^3$ as
the perturbation potential $V_I$
on prong 1. Thus the unperturbed potential is
\beq
V_0(x_i) = {1\over 4}\omega^2 x_i^2~.
\eeq
Following the usual route, the shift of the ground state can be
computed. The first order shift, $\delta E_0$, is given by
\beq
\delta E_0=\int d{x_1} \psi_1^{(0)*}(x_1) V_I \psi_1^{(0)}(x_1)
=\frac{4\alpha}{3\sqrt{2\pi\omega^3}}~.
\eeq

The next energy level
has two fold degeneracy, and hence it is necessary
to use the formalism of degenerate perturbation theory.
We compute the matrix elements of the perturbing potential
$V_I$ using $\psi^{(1)}(\vec{a_1}; \vec{x})$ and
$\psi^{(1)}(\vec{a_2}; \vec{x})$ as basis vectors. For this situation, the
matrix
$V_I$ is explicitly given by
$$V_I =
\left(
\begin{array}{c c}
\gamma               &       {1\over {\sqrt{3}}} 	\gamma\\
{1\over {\sqrt{3}}} \gamma     &       {1\over 3} 	\gamma
\end{array}
\right), $$
where, $\gamma=\frac{8\alpha}{\sqrt{2\pi\omega^3}}$.
Diagonalizing the corresponding Hamiltonian, we find
the eigenvalues to be $0$ and
${4\over 3}\gamma$. Hence, at this level degeneracy is lifted and the
difference in energy is ${4\over 3}\gamma$.
The two eigenfunctions are given by
\beq
\frac{1}{2}\psi^{(1)} (\vec{a_1}; \vec{x}) - \frac{\sqrt{3}}{2}
  \psi^{(1)} (\vec{a_2};\vec{x})=
\frac{1}{\sqrt{2}}\{0,\psi^{(1)} (x_2),- \psi^{(1)} (x_3)\}
\eeq
and
\beq
{1\over 2} \psi^{(1)} (\vec{a_1}; \vec{x}) + {{\sqrt{3} }\over 2}
\psi^{(1)} (\vec{a_2};\vec{x})= \frac{1}{\sqrt{6}}
\left\{ 2\psi^{(1)} (x_1), -\psi^{(1)} (x_2), \psi^{(1)} (x_3)\right\}.
\eeq
It is interesting to note that the first state has the same energy as
the unperturbed system. Thus we see that, since the perturbation
was added to just one prong, it does not break the symmetry completely and
one of the eigenenergies remains unchanged.

\section{Scattering in a Multi-Prong System}

\hspace{.25in}    The scattering of an incident plane wave off the
vertex in a multi-prong system
offers scenarios that are different from the usual
scattering in one-dimensional problems.
As a first example, consider a plane wave with energy $E$ moving
along Prong 1, incident upon the vertex of an $N$-prong
system with a constant potential $V=0$ on all prongs. This is
a trivial example in the two-prong case leading to 100\% transmission,
 and zero reflection.
For more than two prongs, it is not a priori intuitively clear whether
full transmission will occur or not. The wave function on prong 1 is given by
\beq
\psi_1(x_1)=\exp(-ik x_1)+r\exp(ik x_1),
\label{eqn1}
\eeq
where $k$ is the momentum of the incident plane wave $(E=k^2)$.
The wave function on all the other prongs
($i\neq 1$)  only consists of outgoing waves and is given by
\beq
\psi_i(x_i)=t_i \exp(ik x_i).
\label{eqn2}
\eeq
Imposition of the boundary conditions (\ref{BC1}) and (\ref{BC2})
on these wave functions at the vertex
relates the reflected amplitude $r$ to the
transmission amplitudes $t_i$. Their explicit relationship is given
by
\begin{eqnarray}
(1+r)=t_i ~\;\;(i=2,\cdots,N), && r-1+\sum^{N}_{i=2}t_i=0.
\label{scat1}
\end{eqnarray}
The solution is
\beq
r={ {2-N} \over {N} },\;\;\;\; t_i={2\over N} \;\;(i=2,\cdots,N).
\label{coefficients}
\eeq
The reflection and transmission coefficients are then given by
\begin{eqnarray}
{\cal{R}}={{({2-N})^2}\over {N^2}},  &&
{\cal{T}}_i={4\over {N^2}}\;\;\;\;(i=2,\cdots,N). \nonumber
\end{eqnarray}
Clearly, one has ${\cal{R}}+\sum^N_{i=2}{\cal{T}}_i=1$, and the
probability current is conserved. For $N=2$,
the reflection coefficient vanishes, as expected. However, for
$N>2$ there is always a finite amount of reflection. For example, for a
three-prong potential, the reflection coefficient is ${\cal{R}}= {1\over 9}$.
In fact ${\cal{R}}$ increases continuously with $N$, and approaches unity
for large number of prongs, $N$. (This increase of $\cal{R}$ with $N$ is
shown in Fig. 7 by the curve labeled $1$.)
Thus one has the curious result that an
incident wave, when given a large choice of scattering paths, in fact
prefers to be reflected back on its initial prong! \bl

Now let us consider a more general case where
the incident wave is on a prong with potential zero, whereas the
remaining $(N-1)$ prongs are at constant
potential $V_0$. Let the energy of the wave be $E>V_0$. Let us define
$k^2=E$, $k'^2=E-V_0$ and the parameter
$\zeta ={ {k'}\over k}=\sqrt{1-\frac{V_0}{E}}$.
The wave function on prong 1 is again given by
$\psi_1(x_1)=\exp(-ik x_1)+r\exp(ik x_1)$,
and wave functions along remaining $(N-1)$ prongs are given by
$\psi_i(x_i)=t_i \exp(ik' x_i) \;\;\;(i=2,\cdots,N)$.
Boundary conditions (\ref{BC1}) and (\ref{BC2}) applied at the vertex
yield
\beq
(1+r)=t_2=t_3=\cdots=t_N \equiv t \;,\;\;~~-1+r+(N-1)\zeta=0.
\eeq
Solving these equations, we find $r={ {1-\zeta(N-1)}\over {1+\zeta(N-1)}}$
and $t={ 2\over {1+\zeta(N-1)}}$. One can show that $r$ and $t$ obey
$1=r^2+\zeta (N-1) t^2$, as required by conservation of probability.
The reflection coefficient ${\cal{R}}$, which is now given by
$${\cal{R}}=\left[{ {1-\zeta(N-1)}\over {1+\zeta(N-1)}}\right]^2,$$
has a rather interesting behavior as a function of $\zeta$ and $N$.
In Fig. 7, we have plotted
${\cal{R}}$ as a function of $N$ for several values of $\zeta$.
${\cal{R}}$
vanishes for $\zeta ={1\over {(N-1)} }$. Thus for some special values of
${  {V_0}\over E}$, we have complete transmission through the vertex. As $N$
increases, ${\cal{R}}$ slowly approaches unity for all $\zeta$.
 \bl

As yet another interesting example we consider scattering from a T-stub with
$N$ open ended prongs and one prong of finite length $l$ (the stub).
We consider a case where
potentials along all the prongs are zero. A plane wave of momentum $k$
(energy $E=k^2$) incident upon the vertex from prong 1, sets up stationary
waves in the stub, and outgoing plane waves in the remaining $N-1$ prongs.
The wave function on prong 1 is given by $\exp(-ikx_1)+r\exp(ikx_1)$.
Along the stub,  the wave function is given by $A\sin k(l-x_s)$, where $x_s$
is the coordinate along the stub and $A$ is a constant.
Wave functions along all other prongs are given by
$t\exp(ikx_j), \;\;(j=2,\cdots,N)$. From boundary conditions (\ref{BC1})
and (\ref{BC2}), we get:
$$ 1+r=A \sin kl =t\;; \;\;\; ik(r-1)+ik(N -1) t - A k \cos kl =0.$$
Solving them, we get
\begin{equation}
r= { {- (N-2) - {\rm cot}\, kl } \over { N + i\, {\rm cot}\, kl} },
\;\;\;
t= { {2} \over { N + i\, {\rm cot}\, kl} }.
\end{equation}
The reflection coefficient is given by
${\cal R}\equiv |r|^2= { {{\rm cot}\,^2 kl +(N-2)^2 } \over
{{\rm cot}\,^2 kl + N^2 } }$.
Clearly, from the above expression,  one gets
$|r|^2 + (N-1) |t|^2 = 1$. Interestingly, we find that for $kl = n\pi$,
i.e. energy $E={ n^2 \pi^2\over{l^2}}$,  the reflection coefficient
${\cal R}\rightarrow 1$.
At these energies, a standing wave is set up jointly
in the stub and in prong 1, and the particle never enters the other prongs.

Let us generalize the discussion of scattering to an $N$-prong vertex with
identical potentials $V(x_i)$ on all prongs. Define two linearly independent
solutions $f(x_i)$ and $g(x_i)$ of the Schr\"{o}dinger equation using
asymptotic boundary conditions as $x_i \rightarrow \infty$:
\beq
f(x_i)\rightarrow \exp(-ikx_i), \;\;\;g(x_i)\rightarrow \exp(ikx_i).
\eeq
The vertex conditions are
$f(0)\,+\,r\,g(0)=\,t\,g(0)$, and  $f'(0)\,+\,r\,g'(0)+\,(N-1)\,t\,g'(0)=0$.
The solution is
\beq
r = 	-{1 \over N} { {f'(0)} \over {g'(0)}}
	-{{(N-1)} \over N} { {f(0)} \over {g(0)}}, \;\;\;\;
   t = 	-{1 \over N} { {f'(0)} \over {g'(0)}}
	+{ 1 \over N} { {f(0)} \over {g(0)}}.
\label{rt}
\eeq

Conservation of probability requires $|r|^2+(N-1) |t|^2=1$ for all values of
$N$.  This is satisfied provided
$\left| { {f(0)} \over {g(0)}} \right|=
\left| { {f'(0)} \over {g'(0)}} \right|=1.$
Here again, we see that as the number of prongs $N \rightarrow \infty$,
the reflection coefficient ${\cal R}=|r|^2$ approaches unity.
\bl

Having analyzed scattering on a general identical $N$-prong domain,
we now investigate whether we can extract any further information
using supersymmetry.
In one dimensional quantum mechanics, supersymmetry relates
reflection and transmission coefficients of one potential with
those of its supersymmetric partner potential\cite{Akhoury84}.
We find  that this relationship holds also for the case of $N$
identical prongs.
\bl

We have just seen that for the potential $V(x_i)$ on all prongs,
the scattering is described by Eq. (\ref{rt}). The partner potential is
given by Eq. (\ref{V-tilde}). Let $\tilde{f}$ and $\tilde{g}$ be the solutions
of the partner potential which behave like
$\tilde{f}(x_i)\rightarrow \exp(-ikx_i), \;\;
\tilde{g}(x_i)\rightarrow \exp(ikx_i)$, as $x_i\rightarrow \infty$.
Functions $\tilde{f}$ and $\tilde{g}$ are given by
\beq
\tilde{f}(x_i) ={1 \over {W_\infty+ik}}
\left\{  {{-d}\over {dx_i}}  +W(x_i) \right\}
f(x_i),
\;\;
\tilde{g}(x_i) ={1 \over {W_\infty-ik}}
\left\{  {{-d}\over {dx_i}}  +W(x_i) \right\}
g(x_i),
\label{tilde_f}
\eeq
where
$W_{\infty}$ is the value of the superpotential at infinity.
{}From these expressions,  we can show that
${ {\tilde{f}(0)} \over {\tilde{g}(0)}} =
\left( { {W_{\infty}-ik} \over {W_{\infty}+ik}  }\right)
{ {f'(0)} \over {g'(0)}}$, and ${ {\tilde{f}'(0)} \over {\tilde{g}'(0)}} =
\left( { {W_{\infty}-ik} \over {W_{\infty}+ik}  }\right)
{ {f(0)} \over {g(0)}}$.
The reflection and transmission amplitudes for the potential $\tilde{V}(x_i)$
are given by
$$ \tilde{r} = \left( { {W_{\infty}-ik} \over {W_{\infty}+ik}  }\right)
\left(	-{1 \over N} { {f(0)} \over {g(0)}}
	-{{(N-1)} \over N} { {f'(0)} \over {g'(0)}}\right)
, \;\;\;\;
   \tilde{t} = 	\left( { {W_{\infty}-ik} \over {W_{\infty}+ik}  }\right)
\left( -{1 \over N} { {f(0)} \over {g(0)}}
	+{ 1 \over N} { {f'(0)} \over {g'(0)}}\right).
$$

Note, for the special case of two-prongs, these equations agree with
reference \cite{Akhoury84}.
We find the following relationship among $r$, $t$ and their partners
$\tilde{r}$ and $\tilde{t}$:

$$ \tilde{r}=\left( { {W_{\infty}-ik} \over {W_{\infty}+ik}  }\right) r,~~
\tilde{t}=\,- \, \left( { {W_{\infty}-ik} \over {W_{\infty}+ik}  }\right) t.$$
Note that
$|\tilde{r}|^2=|r|^2$, which implies that for the identical prong case,
the reflection coefficients are the same for any potential and its
supersymmetric partner.

Now, using an explicit example, we provide a concrete demonstration of
the usefulness of the machinery that we just developed.
For the sake of simplicity, we shall work with
the example of a free particle in an $N$-prong system for which reflection and
transmission coefficients are given by Eq. (\ref{coefficients}).
The superpotential for this problem is given by $W(x)=\tanh x$.
This superpotential generates two distinct potentials related by
supersymmetry. They are
$V(x) \equiv W^2(x) - W'(x) = 1 - 2\,{\rm sech}^2x$
and
$\tilde{V}(x) \equiv W^2(x) + W'(x) = 1$. The first potential holds
one bound state at energy $E=0$.
\bl

Amplitudes $\tilde{r}$ and $\tilde{t}$ for the free particle system
are ${ {2-N} \over N}$ and $2 \over N$, respectively. Amplitudes
$r$ and $t$ for the potential
$V(x) = 1 - 2\,{\rm sech}^2x$, are then given by
$$ r=   \left( {  {1+ik} \over {1-ik} } \right)
        \left[ -{1 \over N} { { \tilde{f}(0)} \over {\tilde{g}(0)} }
        -{{N-1} \over N} { {\tilde{f}'(0)} \over {\tilde{g}'(0)} }
        \right]
        = \left( {  {1+ik} \over {1-ik} } \right)
        \left( { {N-2} \over {N}  } \right)\;,\;\;
t = -\left( {  {1+ik} \over {1-ik} } \right)
        \left( { {2} \over {N}  } \right)\;.$$

Thus, knowing the reflection and transmission amplitudes $\tilde{r}$
and $\tilde{t}$ for the free particle system, supersymmetry allows us to
determine amplitudes $r$ and $t$ for the rather non-trivial potential
$V(x) = 1 - 2\,{\rm sech}^2x$. Note that coefficients
$\tilde{\cal R}=|\tilde{r}|^2$ and  $\tilde{\cal T}=|\tilde{t}|^2$  are equal
to coefficients for the partner potential,
${\cal R}=|{r}|^2$ and  ${\cal T}=|{t}|^2$ respectively.

\section{Tunneling in a Multi-Prong System}

\hspace{.25in}    Tunneling is another sector of great interest that
shows a marked difference
from the two-prong case. Consider a system with $N$ identical prongs,
each one having one minimum (well-like structure, similar to the two-prong
case shown in Fig. 2). In this section,
we look at the
tunneling of a localized wave packet from one prong to the others. For
simplicity, we start with a three-prong system.  The generalization
to higher number of prongs is then straightforward. Let
$\Phi_i(\vec{x})$ denote a wave packet localized in the $i$-th
prong. Such a packet
can be approximated by the following linear combination of eigenstates
of the system:
\beq
\Phi_i(\vec{x})={1\over 3} \left( \psi^{(0)}(\vec{1},\vec{x}) +
\psi^{(1)}(\vec{a_i},\vec{x}) \right),
\eeq
where $\vec{1}\equiv (1,1,1)$
and $\vec{a_i}$ denotes a three-tuple that has a $2$ in the $i$-th
position and $-1$ in the rest. For example, $\vec{a}_1$ is given by
the three-tuple $(2,-1,-1)$. \bl

The amplitude for the localized state $\Phi_i(\vec{x})$ to tunnel
out of the prong 1 is given by:
\begin{equation}
{\cal{A}}(t)=\left \langle \psi_f \right|\exp(-iHt)\left| \psi_i
\right \rangle
\label{tunnel1}
\end{equation}
where
\begin{eqnarray}
\psi_i=\Phi_1(\vec{x})  &  {\rm and}  &  \psi_f={1\over
{\sqrt{\alpha_2^2+\alpha_3^2}}} \left(\alpha_2 \Phi_2 (\vec{x}) + \alpha_3
\Phi_3 (\vec{x}) \right); \nonumber
\end{eqnarray}
${\alpha_2+\alpha_3=1}$. Substituting explicit expressions for
$\psi_f$ and $\psi_i$ in Eq. (\ref{tunnel1}), we get
\begin{eqnarray}
{\cal{A}}_{\alpha_2\,\alpha_3} &=& {1\over {3\sqrt{\alpha_2^2+\alpha_3^2}}}
\int d\vec{x}\, \left[  \alpha_2 \Phi_2 (\vec{x}) + \alpha_3 \Phi_3
(\vec{x}) \right] \left[ e^{-iE_0\,t} \psi^{(0)}(\vec{1},\vec{x}) +
e^{-iE_1\,t} \psi^{(1)}(\vec{a_1},\vec{x}) \right] \nonumber \\
&=& {1\over {9\sqrt{\alpha_2^2+\alpha_3^2}}}
\int d\vec{x}\, \left[
\alpha_2 \psi^{(1)}(\vec{a_2},\vec{x}) +
\alpha_3 \psi^{(1)} (\vec{a_3},\vec{x}) +
\psi^{(0)} (\vec{1},\vec{x}) \right] \nonumber \\
&& \times \left[ e^{-iE_0\,t} \psi^{(0)}(\vec{1},\vec{x}) +
e^{-iE_1\,t} \psi^{(1)}(\vec{a_1},\vec{x}) \right] \nonumber \\
&=& {{e^{-iE_1\,t}}\over {9\sqrt{\alpha_2^2+\alpha_3^2}}}
\left[ \alpha_2 \vec{a_1} \cdot \vec{a_2} + \alpha_3 \vec{a_1} \cdot
\vec{a_3} + 3\,e^{i\delta E\,t} \right],
{}~~~\delta E \equiv E_1-E_0~,
\nonumber
\end{eqnarray}
where we have used $\int d\vec{x}\, \psi^{(1)}(\vec{a_1},\vec{x})
\psi^{(1)}(\vec{a_2},\vec{x}) = \vec{a_1} \cdot \vec{a_2}=
\sum_i (\vec{a_1})_i (\vec{a_2})_i $. Substituting
$\vec{a_1} \cdot \vec{a_3}=\vec{a_1} \cdot \vec{a_2}=-3$,
we get
$${\cal{A}}_{\alpha_2\,\alpha_3} =
{{e^{-iE_1\,t}}\over {3\sqrt{\alpha_2^2+\alpha_3^2}}}
\left[ -1 + e^{i\delta E\,t} \right].$$
The probability of tunneling is then given by
$${\cal{P}}_{\alpha_2\,\alpha_3} = \left|
{\cal{A}}_{\alpha_2\,\alpha_3}\right|^2
={ {4} \over {9(\alpha_2^2+\alpha_3^2)} } \sin^2
\left({ {\delta E\,t} \over 2} \right);$$
The maximum value for this tunneling probability is $8\over 9$, and
that occurs when $\alpha_2=\alpha_3={1\over 2}$. It is worth noting here
that unlike the two-prong case, the wave packet never completely
goes out of the prong 1.

When we generalize this situation to $N$-prong case, we find that the
probability amplitude for a wave packet $\Phi_1(\vec{x})$
of tunneling out of prong 1
is given by
$${\cal{A}}_{\vec{\alpha}} =
{
	{2ie^{-i(E_0+\delta E/2) } {\sin (\delta E t/2)} }
	\over
	{ N \left(\sum_{j=2}^N \alpha_j^2 \right)^{1/2} }
}~~,$$
where ${\vec{\alpha}}\equiv \{\alpha_2, \alpha_3, \cdots, \alpha_N \}$ with
${\alpha_i} \over \left(\sum_{j=2}^N \alpha_j^2 \right)$ giving the
amplitude for finding the packet in the $i$-th prong
$\left(
{
{\sum_i \alpha_i \Phi_i(\vec{x})}
\over
{\left(\sum_{j=2}^N \alpha_j^2 \right)} }
\right)$. This amplitude is
maximum for the symmetric case for which all ${\alpha_i} $ are equal.
In this case the probability of tunneling is given by
$${\cal{P}}=
{ {4(N-1)} \over N^2}
\sin^2 \left({ {\delta E\,t} \over 2} \right).$$
Thus the probability of tunneling decreases when more alternatives are
available, and goes to zero as the number of prongs becomes very large.
This is similar to the scattering situation discussed in Sec. 7. \bl \bl

{\bf Acknowledgements:}
It is a real pleasure for one of us (U.S.) to thank Dr. Arun Jayannavar
for helpful initial discussions.
This work was supported in part by the
U. S. Department of Energy under grant DE-FGO2-84ER40173.

\bl

\newpage
{\large {\bf Figure Captions} } \bl\bl

\noi {\bf Figure 1:}
 Schematic diagram of an $N$-prong potential. The position on any
prong $i$ is specified by the positive coordinate $x_i$, with $x_i=0$ being
the vertex. The angles between prongs play no role in any computations. \bl

\noi {\bf Figure 2:}
Example of a potential with two identical prongs. The positive
variables $x_1,x_2$ together span the real $x$-axis $(-\infty < x < \infty)$.
\bl

\noi {\bf Figure 3:}
The three lowest eigenfunctions [Eqs.
(\ref{wave_ho}), (\ref{wave_ho2}) and (\ref{excited_ho})]
for a three-prong harmonic oscillator
potential $V(x_i)={1 \over 2}m\omega^2x_i^2$. The ground state $\psi^{(0)}$
is non-degenerate, whereas the first excited state $\psi^{(1)}$ is
doubly degenerate. \bl

\noi {\bf Figure 4:}
The three lowest energy eigenstates (un-normalized) for a
three-prong free particle potential. The solid lines correspond to
$l_1=l_3=1,~l_2=0.8~$. The dashed lines correspond to the identical prong case
$l_1=l_2=l_3=1~$. \bl

\noi {\bf Figure 5:}
The eigenvalues $k$ (energy $E=k^2$) for a three-prong free
particle potential as a function of $l_2$, the length of prong 2. The
lengths of the other two prongs are kept fixed and {\it equal} $l_1=l_3=1~$.
Note the pattern of degeneracy and level crossings. The curves are
labeled by the number of nodes in the wave function (in all prongs). \bl

\noi {\bf Figure 6:}
The eigenvalues $k$ (energy $E=k^2$) for a three-prong free
particle potential as a function of $l_2$, the length of prong 2. The
lengths of the other two prongs are kept fixed and {\it unequal} $l_1=1~$,
$l_3=\sqrt {2}$. There is now no degeneracy. The curves are labeled by
the number of nodes in the wave function (in all prongs). \bl

\noi {\bf Figure 7:}
A plot of the reflection coefficient $\cal{R}$ versus the
number of prongs $N$. The incident wave with energy $E$ is on prong 1 with
zero potential. All other prongs are taken to be at a constant potential
$V_0$. The curves are labeled by the parameter $\zeta
\equiv \sqrt {1-\frac{V_0}{E}}$.

\end{document}